\documentclass[prd]{revtex4}

\usepackage[english]{babel}

\usepackage[latin1]{inputenc}

\usepackage{amsmath, amsfonts,amssymb}

\usepackage{amstext}

\usepackage[colorlinks, citecolor=blue, backref]{hyperref}

\providecommand{\abs}[1]{\lvert#1\rvert}

\providecommand{\Lie}{\pounds}

\def\<<{``}

\def\D{\partial}

\def\ga{\gamma}

\def\al{\alpha}

\def\B{\beta}

\def\Si{\Sigma}

\def\la{\lambda}

\def\de{\delta}

\def\/{\, /\;}

\begin{document}

\title{The relative energy of homogeneous and isotropic universes from variational principles\footnote{
This paper is published despite the effects of the Italian law 133/08. 
        This law drastically reduces public funds to public Italian universities, which is particularly dangerous for scientific free research, 
        and it will prevent young researchers from getting a position, either temporary or tenured, in Italy.
        The authors are protesting against this law to obtain its cancellation.\\
         See http://groups.google.it/group/scienceaction\\}}
        
\author{Enrico Bibbona}
\email{enrico.bibbona@unito.it}
\affiliation{Department of Mathematics, University of Torino,\\
Via Carlo Alberto 10, 10123 Torino, Italy\\
INFN, Sezione di Torino, Iniziativa Specifica NA12 (Italy)}
\author{Lorenzo Fatibene}
\email{lorenzo.fatibene@unito.it}
\affiliation{Department of Mathematics, University of Torino,\\
Via Carlo Alberto 10, 10123 Torino, Italy\\
INFN, Sezione di Torino, Iniziativa Specifica NA12 (Italy)}
\author{Mauro Francaviglia}
\email{mauro.francaviglia@unito.it}
\affiliation{Department of Mathematics, University of Torino,\\
Via Carlo Alberto 10, 10123 Torino, Italy\\
INFN, Sezione di Torino, Iniziativa Specifica NA12 (Italy)\\
LCS, Universit\`a della Calabria (Italy)}

\date{\today}

\begin{abstract}

\noindent We calculate the relative conserved currents, superpotentials and conserved quantities between two homogeneous and isotropic universes. In particular we prove that their relative \<<energy'' (defined as the conserved quantity associated to cosmic time coordinate translations for a comoving observer) is vanishing and so are the other conserved quantities related to a Lie subalgebra of vector fields isomorphic to the Poincar\'e algebra. These quantities are also conserved in time. We also find a relative conserved quantity for such a kind of solutions which is conserved in time though non-vanishing.
This example provides at least two insights in the theory of conserved quantities in General Relativity. First, the contribution of the cosmological matter fluid to the conserved quantities is carefully studied and proved to be vanishing. Second, we explicitly show that our superpotential (that happens to coincide with the so-called KBL potential although it is generated differently) provides strong conservation laws under much weaker hypotheses than the ones usually required. In particular, the symmetry generator is not needed to be Killing (nor Killing of the background, nor asymptotically Killing), the prescription is quasi-local and it works fine in a finite region too and no matching condition on the boundary is required.

\end{abstract} 
\maketitle

\section{Introduction}
The definition of energy is one of the most disputed topics in General Relativity.
In the case of isolated systems, such as black holes, different prescriptions are available to define energy (as well as for the other conserved quantities) and on  the simplest known examples where most of them are applicable, they usually provide the same result \cite{sbados}.
In these prescriptions at least one of the following requirements is always asked for: the symmetry generators are Killing vectors (at least of the background, or at least asymptotically); some matching between the dynamical metric and the background; some special asymptotic behavior.
In the case of cosmological solutions, as we shall see, all these techniques cannot be applied and there is no general agreement not only on how much energy is there (nor in the universe nor in its local domains), but we do not even have a clear definition of what the energy should be. 
There are References (see for example Refs. \cite{rosen, berman, vargas}) where the energy of some models of homogeneous and isotropic universes is defined according to different prescriptions (mainly relying on pseudotensors) and it turns out to be vanishing; however some criticism to these non-covariant methods can be found for example in Ref \cite{cooperstock} and General Relativity principles are quite clear on this: either such quantities can be described in some covariant way or they are not fundamental to the description of the Universe. 
Recently three articles have been published with definitions leading to different results (see. Refs \cite{chen2, garecki2, garecki_new}).

From the purely physical viewpoint the absence of a region far off from the sources makes very hard to give a physical interpretation to conserved quantities that however can be calculated on a purely mathematical ground under very general hypothesis once one clearly fixes the definitions.

In recent years many authors (\cite{chen,katz,wald,julia,reggeteitelboim}) provided prescriptions to compute the conserved currents in General Relativity based on a suitable \<<boundary correction'' to the Komar superpotential relative to a reference configuration.

To this research effort two of the present authors gave a number of contributions (see for example Refs.\cite{cavalese, grg, wecriticize, brownyork, boundary}) mainly proposing a pure variational route to the definition of the relative conserved currents between two solutions of the field equations interpreted as the {\em amount} of conserved current needed to pass from one solution to the other. The mathematical framework introduced to this aim goes well beyond General Relativity and extends in fact to any gauge-natural Lagrangian field theory (see for example Ref. \cite{lo} and Ref. \cite{augmented}).

The aim of this paper is to apply and adapt the previously known techniques to the calculation of the relative conserved quantities in the case of two isotropic cosmological solutions with different densities, where some technical requirements commonly found in the literature are not fulfilled and where the contribution of matter has to be carefully taken into account.

The physical motivation behind this choice is to understand how much a change of the scale factor (and thence of the matter amount of the Universe) affects the conserved currents and in particular the gravitational energy. 

We find that all the relative covariantly conserved quantities related to a Lie subalgebra of the Lie algebra of vector fields, isomorphic to the Lie algebra of the Poincar\'e group, are vanishing and in particular this holds true for the relative energy that we define as the conserved quantity associated to cosmic time translations for a comoving observer. Let us remark that these observers are not chosen a priori (this would not be allowed by the covariance principles of General Relativity) but they are a posteriori selected by this family of cosmological solutions as {\em preferred cosmological observers} who provide a better interpretation of the results. In principle our prescription applies to {\em any} observer.
According to our interpretation we can say that \<<injecting'' matter into the Universe do not change its conserved quantities.

\section{The variational framework for relativistic hydrodynamics}

Let us review here a variational formulation of relativistic hydrodynamics that was originally presented in detail in Ref. \cite{en1}. It is in practice equivalent to the ones of Refs. \cite{taub, kijowski-art, he} but it shows some formal and computational advantages. The aim of reviewing it here is to prove (and not just assume as it is usually done) that the matter Lagrangian does not contribute to the conserved quantities related to the diffeomorphism invariance of the theory.

Our kinematical description is given trough the following fields:
gravitation is represented by a Lorentzian metric with components $g_{\mu\nu}$, while the fluid 
degrees of freedom are encoded in a nowhere vanishing vector density $J=J^\mu ds_\mu$ (where $ds_{\mu}$ is the natural basis of $(m-1)$-forms over an $m$-dimantional manifold. The form $J$ is such that 
the unit timelike vector
$u^\mu=J^\mu /\abs{J}$ is tangent to the flow lines and that 
the volume form
$\sqrt{\abs{g}}\rho=\abs{J}$ represents the matter density. Here $g$ denotes the determinant of the metric tensor.

In order to implement conservation of matter we only 
allow for closed $J$ (i.e. $\D_\mu J^\mu=0\equiv\nabla_\mu J^\mu$), so that for any
closed  $3$-surface $\Si$ in spacetime the flow of $J$ through $\Si$ vanishes.

Starting from the Lagrangian \begin{equation}
L = L_{H} + L_F \label{L1}\end{equation} with

\[\begin{aligned}
{}&L_H ( g_{\al\B}, \D_\la g_{\al\B}, \D_{\la\omega} g_{\al\B}) = \frac{\sqrt{\abs{g}}}{2\kappa}\: (R-2\Lambda) \, ds\\
& L_F (J^\mu, g_{\al\B}) = -\sqrt{\abs{g}}\: [\rho\;(1+ e (\rho))]\, ds = -\sqrt{\abs{g}}\: \mu (\rho)\, ds 
\end{aligned}\]
where $R$ is the scalar curvature of the metric, $Lambda$ a constant and $e (\rho)$ a 
suitable function (here generic) of the scalar
$\rho= \sqrt{\frac{g_{\mu\nu}J^\mu J^\nu}{\abs{g}}}$, physically 
interpreted as the internal energy of the fluid. Pressure is defined as $P= \rho^2 \frac{\D e}{\D \rho}$.

We want to minimize the action functional preserving the constraint $\D_\al J^\al=0$.
To do so we take arbitrary variations for the metric, while $J$ is varied according to the rule  $\de J^\mu =\Lie_X J^\mu = \D_\nu
J^\mu X^\nu - J^\nu \D_\nu X^\mu + J^\mu\D_\nu X^\nu$. Intuitively, it amounts to drag the flow lines of the fluid in the spacetime along an arbitrary diffeomorphism (of course fixing boundary values). This euristic prescription can be formally justified on rigorous bases; see Ref. \cite{en1}.

The Euler-Lagrange equations arising from this \<<constrained'' prescription are the following 

\begin{equation}\left\{\begin{aligned}{}&\nabla_\mu J^\mu=0\qquad\quad\text{(constraint)}\\
&R_{\mu\nu} -  \frac{1}{2} (R-2\Lambda) g_{\mu\nu} = \kappa H_{\mu\nu}\\
&(u^\cdot_\mu u^\nu
+ \de^\nu_\mu)  \nabla_\nu P + ( \mu + P )\: u^\nu \nabla_\nu 
u^\cdot_\mu =0,
\end{aligned}\right.\label{eqns}\end{equation}
where $u^\mu=J^\mu /\abs{J}$ and $H_{\mu\nu}$ is the fluid stress tensor
\[H_{\mu\nu}= P g_{\mu\nu} +(\mu + P) u^\cdot_\mu u^\cdot_\nu.\]

Diffeomorphism covariance of the Lagrangian \eqref{L1} provides the following N\"other current associated to any spacetime vector field $\xi=\xi^\nu \D_\nu$
\begin{equation}\label{corrente1}\mathcal{E}(L,\xi)=\mathcal{E}^\al(L,\xi) \, ds_\al = \mathcal{E}_{H}(L_{H},\xi)^\al \, ds_\al - \sqrt{\abs{g}}\,(g^{\al\mu}\, H_{\mu\nu} \xi^\nu)\,ds_\al \end{equation} 
where we set
\begin{equation}
\label{corrente2}\mathcal{E}^\al_H(L_{H},\xi) = \frac{\sqrt{\abs{g}}}{2\kappa}\left[\left(
\frac{3}{2}R^\al_{\cdot\la} -(R-2\Lambda)\de^\al_\la\right) \xi^\la
\left( g^{\B\gamma}\de^\al_\la - g^{\al(\ga}\;\de^{\B)}_\la \right) \nabla_{\B\ga} \xi^\la\right] 
\end{equation}
for the N\"other current obtained from the Hilbert Lagrangian $L_H$.
The current $\mathcal{E}(L,\xi)$ uniquely splits according to the general theory on gauge-natural field theories (see \cite{lo}, chapter 6) into $\mathcal{E}(L,\xi)=\tilde{\mathcal{E}}(L,\xi)+\text{Div }\mathcal{U}(L,\xi)$ where the reduced current $\tilde{\mathcal{E}}(L,\xi)$ vanishes on shell (i.e. when computed along a solution of the field equations)  while the superpotential $\mathcal{U}(L,\xi)$ has the usual Komar expression
\begin{equation}
\mathcal{U^{\al\mu}}(L,\xi) =\frac{1}{2\kappa} \nabla^{[\mu} \xi^{\al]}. \label{komar}\end{equation} The fluid Lagrangian does not contribute to the superpotential since its contribution to the N\"other current is of order zero in $\xi$ (see again Ref. \cite{en1} and the general theory in Ref. \cite{lo} for more details).

\section{Relative conservation laws for homogeneous and isotropic universes and relativistic hydrodynamics}

A homogeneous and isotropic cosmological model is a particular class of solutions of the system described by the Lagrangian \eqref{L1} where the couple $(g,J)$ is such that in a system of synchronous comoving coordinates $(t, r, \theta, \phi)$ it looks locally like (see for example \cite{gravitation})
\[ g= -dt^2 +a^2(t)\left[\frac{dr^2}{1-K r^2} + r^2 \left( d\theta^2 +\sin^2 \theta\, d\phi^2 \right) \right] \] and
\[J=\sqrt{\abs{g}}\rho\, ds_0,\]
where $ds_0= \D_t \lrcorner ds$ and $K\in \{-1,0,1\}$ is the normalized constant scalar curvature of the spatial leafs at constant $t$, while the scale factor $a(t)$ and the density $\rho$ are subjected to equations \eqref{eqns} that in the chosen coordinates read as
\begin{equation}\left\{\begin{aligned}{}&
\D_t (\sqrt{\abs{g}}\,\rho)=0\\
&3\dot{a}^2 + 3k - \kappa \mu a^2 -\Lambda a^2=0\label{equazioni}\end{aligned}\right.\end{equation}
being in this case the third equation in \eqref{eqns} equivalent to the first one. 

Quasi-local conserved quantities in a closed region $D$ of spacetime for a given solution of the field equations and relative to a Lagrangian symmetry can be defined as the integral of the superpotential along the boundary of $D$. This definition however is affected by a number of ambiguities that have been thoroughly discussed in the literature (see for example Ref. \cite{lo} Chapter 6, Section 5 and Ref. \cite{wecriticize}).

A better defined concept is that of {\it relative} conserved currents and quantities. They arise in any gauge-natural Lagrangian field theory (of any order) from the following prescription. Let be $y^{i}$ be the fields, $y^{i}_{\mu}$ their derivatives (first order for simplicity, but higher order are allowed too) and $L(x^{\mu}, y^{i}, y^{i}_{\mu} )$ a gauge-natural Lagrangian (any Lagrangian that is covariant under spacetime diffeomorphisms or gauge transformations belongs to this class, cf. Ref. \cite{lo}) with first variation formula $\de_{Y} L =E(L,Y)+ \text{Div }F(L,Y)$ where $E(L,Y)=0$ are the field equations and $F(L,Y)$ is the so-called Poincar\'e-Cartan morphism.
Let us consider a one parameter family $\{y^{i}_{s}(x^{\mu})\}_{s\in [0,1]}$ of solutions of the field equations such that $y^{i}_{0}=\bar{y}^{i}$ is interpreted as a reference background and $y^{i}_{1}=y^{i}$ is a solution representing the physical fields.
Our prescription amounts to correct the superpotential in order to fix the zero level for the conserved quantities in such a way that they become vanishing when the physical fields coincide with the given reference. To this aim, however, covariant-ADM formalism (and also other derivations, see Refs.\cite{covadm1, covadm2, brownyork, fromeqns} ) suggests to consider a new {\it relative} superpotential $ \mathcal{U}^{c}(\xi,y_{s}, \bar{y})$ built from the difference of the values of the old superpotential along $y_{s}^{i}$ and $\bar{y}^{i}$, corrected  in such a way that its variation along the infinitesimal generator $X$ of the $1$-parameter family $\{y_{s}(x^{\mu})\}_{s\in [0,1]}$ gives

\begin{equation} \de_X   \mathcal{U}^{c}(\xi,y_s, \bar{y}) =\de_X   \mathcal{U}(\xi, \bar{y}) + i_\xi F(L,X).\label{variation}\end{equation}

In vacuum General Relativity a superpotential the variation of which satisfies \eqref{variation} is known to be the following one

\begin{equation}
\mathcal{U}^{c}_{H}(\xi,g, \bar{g})
= \sqrt{\abs{g}} \nabla^{[\al} \xi^{\B]}-\sqrt{\abs{\bar{g}}}\bar{\nabla}^{[\al} \xi^{\B]} + \sqrt{\abs{\bar{g}}} \bar{g}^{\mu\nu} w^{[\B}_{\mu\nu} \xi^{\al]} ds_{\al\B}.\label{relsup}
\end{equation}
where $g$ is interpreted as the physical metric while $\bar{g}$ is the reference configuration (sometimes called a dynamical background as it is required to be a solution of Einstein equations) and $w^{\mu}_{\al\B}=\Gamma^{\mu}_{\al\B}- \de^\mu_{(\al}\Gamma^\sigma_{\B)\sigma}- \bar{\Gamma}^{\mu}_{\al\B}- \de^\mu_{(\al}\bar{\Gamma}^\sigma_{\B)\sigma}$. Barred quantities are defined with respect to the background metric.

This relative superpotential, as claimed, is the difference between the two copies of the Komar superpotential \eqref{komar} evaluated on $g$ and $\bar{g}$ plus the covariant-ADM correction and it vanishes when $g=\bar{g}$. This superpotential happens to coincide with the so called KBL superpotential (named after the authors of Ref. \cite{katz}) that, however, was derived in a different ad-hoc way and under stricter hypothesis rather than being shown to be the specific case of a much wide concept.

We define now
\begin{equation}
Q(\xi,D,g,\bar{g})= \int_{\D D} \mathcal{U}^{c}(\xi,g, \bar{g}).\label{Q}
\end{equation} 
as the {\it relative conserved quantities} between the two (homotopical) solutions $g$ and $\bar{g}$ in the domain $D$ and relatively to the Lagrangian symmetry generator $\xi$. In other words $Q$ may be interpreted as the {\it cost} in terms of the conserved quantity needed to pass from the solution $\bar{g}$ to the solution $g$ (for any fixed region and for any vector field) .

We point out that besides many side conditions are usually required in the literature in order to give a precise physical interpretation to the conserved quantities ($\xi$ is often asked to be a Killing vector either of $g$ or $\bar{g}$, $D$ is the spatial infinity, $g$ and $\bar{g}$ are usually required to match on $\D D$), since the Hilbert Lagrangian is invariant under diffeomorphisms, the relative superpotential \eqref{relsup} is a strongly conserved current associated to any vector field $\xi$ without any restriction on $\xi$ itself, on $D$, and also without boundary conditions on $g$ and $\bar{g}$. In Appendix \ref{intime} we review a further theoretical distinction between general conserved quantities wit respect to those that are also conserved in the time induced by an ADM foliation of spacetime.

In presence of a cosmological fluid we already noticed that the matter contribution to the Komar superpotential is vanishing. Let us however explore the need for some covariant-ADM-like corrective boundary term such as that appearing in the relative superpotential \eqref{relsup} when one introduces as a reference background another FRW universe with a different amount of matter (thus different scale factor), but same topology for the spacial leaves. The choice of such a background is motivated by the idea of understanding how much a change of the scale factor (and thence of the matter amount of the Universe) affects the conserved currents and in particular the gravitational energy.

Such a corrective term  needs to satisfy condition \eqref{variation} for the fluid Lagrangian that reads
\[  \de_X   \mathcal{U}^{F}(\xi,g_s, \bar{g}) =\de_X  i_\xi F(L_F,X) \]
with $X$ the generator $X = (X_g , X_J)$ of the $1$-parameter family of solutions $(g^{(s)}_{\mu\nu},J_{(s)}^\al)$ with $(g^{(1)}_{\mu\nu},J_{(1)}^\al)$ being the physical configuration (a FRW universe with or without cosmological constant) and $(g^{(0)}_{\mu\nu},J_{(0)}^\al)$ the reference one (another FRW universe with a different amount of matter, but same topology) and where $F(L_F,X)$ is the term that goes under divergence in the first variation of the fluid Lagrangian (the so-called Poincar\'e-Cartan morphism) taken with respect to $X$.

In order to find such a boundary correction to the superpotential let us notice that in a homogeneous and isotropic cosmological model the unique parameter left free by the symmetry requirements is the initial condition for the second of equations \eqref{equazioni} and thence a $1$-parameter family $(g_s, J_s)$ of homogeneous and isotropic universes is allowed to depend on the parameter just through the expansion factors $a_s(t)$. Let us denote from now on by $a(t)$ and $\bar{a}(t)$ respectively the expansion factors of $g$ and of $\bar{g}$. Due to the first of equations \eqref{equazioni}, the $J_s$ cannot depend on $a_s(t)$ and thence they are all the same  $J_s=J$ for each $s$. The second component $X_J$ of the generator of the family is henceforth vanishing. As the fluid Lagrangian does not involve derivatives of $g$, the candidate corrective term is necessarily vanishing.

In the case of homogeneous and isotropic cosmological solution, thence, the superpotential associated to any infinitesimal generator $\xi$ of diffomorphisms has still the same expression as that of equation \eqref{relsup}.

No restriction on $\xi$, however, is required for the conservation law $\nabla_{\mu} \mathcal{E}^{\mu}=0$ to hold on shell, nor for the on-shell identity $\mathcal{E}^{\mu}= \nabla_{\nu} \mathcal{U}^{[\nu\mu]}$ and thence nor for the the strong (off-shell) conservation of $\nabla_{\nu} \mathcal{U}^{[\mu\nu]}$.

In relation with the theory proposed in Ref. \cite{augmented} we remark that the superpotential for the isotropic cosmologies can be directly derived from the augmented Lagrangian

\begin{equation}
\label{aug_lag_testo}L_{HF\text{aug}}= L_H (g)-L_H (\bar{g}) -d_\al (\sqrt{\abs{\bar{g}}} \bar{g}^{\mu\nu} w^\al_{\mu\nu} )
+ L_F(J,g) - L_F(\bar{J}, \bar{g}) 
\end{equation}

This Lagrangian is invariant under diffeomorphisms. To any vector field $\xi$ we can associate the relative N\"other current $\mathcal{E}_{\text{rel}}(L_{HF\text{aug}},\xi, (g,J),(\bar{g},J))=\mathcal{E}_{HF\text{rel}}^\al ds_\al$ where 
\begin{equation} \mathcal{E}_{HF\text{rel}}^\al=\mathcal{E}_{HF}^\al (g, J) - \mathcal{E}_{HF}^\al (\bar{g}, J) + d_\B \left(\sqrt{\abs{\bar{g}}} \bar{g}^{\mu\nu} w^{[\al}_{\mu\nu} \xi^{\B]}\right)\label{corrente_testo} \end{equation}
and $\mathcal{E}_{HF}^\al (g, J)$ and $\mathcal{E}_{HF}^\al (\bar{g}, J)$ can be calculated according to formula \eqref{corrente2}.

Superpotential \eqref{relsup} arises from the splitting of the current \eqref{corrente_testo} according to the general theory (see again Ref.\cite{lo}, chapter 6).

\section{The vanishing of the relative energy and discussions on the other conserved quantities}

Even if the definition of energy is not an undisputed topic, we already said that in many cases it has been fruitfully defined as the flow of the conserved current associated to the generator of time translations through a surface $\Sigma$ at $t=\text{const}$ delimited by a 2-dimensional boundary $\D \Sigma$. Cosmological solutions carry a preferred notion of time to be used to this aim.

Let us remark that if $t$ is the cosmic time, $\D_{t}$ is not a Killing vector for $g$ nor for $\bar{g}$, thence these metrics are not invariant for cosmic time translations. However $\D_{t}$ as any other generator of diffeomorphisms is a Lagrangian symmetry for \eqref{L1} and this allow to calculate the strongly conserved current and quantity related to it by N\"other theorem.

To call the above quantity {\it energy} is now partially a matter of interpretation (and thus questionable in absence of a clear physical definition for this kind of spacetime); however, at least from the mathematical viewpoint, this is a well defined covariantly conserved current associated to a clearly distinguished time and we cannot see any strong argument against its interpretation as a kind of energy. Moreover, this is a trivial extension of the same techniques which have proven to be effective in the cases where the physical interpretation allows for a clear identification of what should be understood as energy on a physical instance (cf. \cite{fatibenemann, btz, brownyork,brownyorkoriginal, wecriticize, taubbolt, raiterimann}).

Let us thence compute the relative energy between two homotopic  homogeneous and isotropic universes (thus two universes with the same $K$, no matter if it is $-1, 0$ or $1$, but different scales $a(t)$ and $\bar{a}(t)$) as the conserved quantity $Q(L_{\text{aug}},\D_{t},\Sigma,(g,J), (\bar{g},J))$ with respect to a hypersurface $\Sigma$.

Straightforward computations (here performed with the aid of Maple tensor package) lead us to conclude that the N\"other  current $\mathcal{E}=\mathcal{E}^{\mu} ds_{\mu}$  \eqref{corrente_testo} has the first component vanishing on-shell, while the other components are identically vanishing.
The superpotential is identically vanishing (in agreement with general theorems on the uniqueness of the splitting) and in particular all of its three summands vanish independently (the relative correction plays no role).
Whichever is the surface $\Sigma$ we conclude that the relative energy contained inside $\D \Sigma$ is vanishing. From the physical viewpoint this means that not only the total energy of the universe as a whole vanishes, but also locally (quasi-locally) in any 3-dimensional domain bounded by a 2-surface $\D \Sigma$ there is no total energy, no matter how much matter is there contained. 
A possible physical interpretation of this fact is that the internal energy of the cosmological fluid increases in fact with the density and it fuels the expansion, while the energy of the gravitational field increases with the density too but it fuels contraction (being a bonding interaction it contributes with the minus sign) and the two contributions balance giving rise to a vanishing total energy.
Let us remark that the energy (being independent on $\Sigma$) besides being a covariantly conserved quantity is also in this case conserved {\it in time}.

Let us now consider the vector fields $T_x, T_y ,T_z$ defined as follows
\begin{align*}T_{x}&= \sin\theta\cos\phi \frac{\partial}{\partial r}+\frac{\cos\theta\cos\phi}{r}\frac{\partial}{\partial \theta}-\frac{\sin\phi}{r\sin\theta}\frac{\partial}{\partial \phi}\\
T_{y}&=\sin\theta\sin\phi \frac{\partial}{\partial r}+\frac{\cos\theta\sin\phi}{r}\frac{\partial}{\partial \theta}+\frac{\cos\phi}{r\sin\theta}\frac{\partial}{\partial \phi}\\
T_{z}&=\cos\theta \frac{\partial}{\partial r}-\frac{\sin\theta}{r}\frac{\partial}{\partial \theta}.
\end{align*}
These vectors are the generators of translations along the following cartesian-like coordinates $x= r\sin(\theta)\cos(\phi)$, $y=r \sin(\theta)\sin(\phi)$, $z=r\cos(\theta)$.
Let moreover $\D\Sigma_{t_0,r_0}$ be the 2-dimensional hypersurface defined by $r=r_0$ and $t=t_0$.
The conserved quantities
\begin{align*}{}&Q(L_{\text{aug}},T_{x},\Sigma_{(t_0,r_0)},(g,J), (\bar{g},J))=\\
&=Q(L_{\text{aug}},T_{y},\Sigma_{(t_0,r_0)},(g,J), (\bar{g},J))=\\
&=Q(L_{\text{aug}},T_{z},\Sigma_{(t_0,r_0)},(g,J), (\bar{g},J))=0\end{align*}
are vanishing (and thus also conserved in time according to the definition of Appendix \ref{intime}).
Let us however notice that in these cases the superpotentials are no longer vanishing and a different choice of the surfaces on which to compute the integral may lead to a different result.

For the vector fields $L_x= y T_z -z T_y$,  $L_y= z T_x -x T_z$ and $L_z=xT_y-yT_x$ that generate rotations of the hypersurphaces of homogeneity we also have
\begin{align*}{}&Q(L_{\text{aug}},L_{x},\Sigma_{(t_0,r_0)},(g,J), (\bar{g},J))=\\
&=Q(L_{\text{aug}},L_{y},\Sigma_{(t_0,r_0)},(g,J), (\bar{g},J))=\\
&=Q(L_{\text{aug}},L_{z},\Sigma_{(t_0,r_0)},(g,J), (\bar{g},J))=0\end{align*}
again without an identically vanishing superpotential.

If we now consider the vector fields $B_x= x \D_t +t T_x$,  $B_y= y \D_t -t T_y$ and $B_z=z\D_t-tT_z$ that generate boosts in the coordinates $(x,y,z)$ defined above we find
\begin{align*}{}&Q(L_{\text{aug}},B_{x},\Sigma_{(t_0,r_0)},(g,J), (\bar{g},J))=\\
&=Q(L_{\text{aug}},B_{y},\Sigma_{(t_0,r_0)},(g,J), (\bar{g},J))=\\
&=Q(L_{\text{aug}},B_{z},\Sigma_{(t_0,r_0)},(g,J), (\bar{g},J))=0\end{align*}
again without an identically vanishing superpotential.

The subalgebra of the Lie algebra of vector fields generated by the ten vector fields $\D_t$, $T_x$, $T_y$, $T_z$, $L_x$, $L_y$, $L_z$, $B_x$, $B_y$, $B_z$ is isomorphic to the Lie algebra of the Poincar\'e group. The covariantly conserved quantities associated to them inside a surface $\D\Sigma_{t_0,r_0}$ defined by $r=r_0$ and $t=t_0$ are also conserved in time and vanishing.

A natural question then arises also in consideration of the analysis carried out in a very famous paper on vacuum fluctuations (see Ref. \cite{tryon}): is it possible to find a non-vanishing covariantly conserved quantity inside $\D\Sigma_{t_0,r_0}$, possibly also conserved in time? The answer is yes. If we compute the conserved quantity associated to the vector field $Z=\frac{\text{arctan}(r)}{a(t)+\bar{a}(t)}\D_t$ for an homogeneous and isotropic universe with negative spatial curvature we find 
\[Q(L_{\text{aug}},Z,\Sigma_{(t_0,r_0)},(g,J), (\bar{g},J))=\frac{r^2}{4\sqrt{1+r^2}}\]
that is also conserved in time (according to the definition of Appendix \ref{intime}).

Let us notice that homogeneous and isotropic universes are quite different from the other cases where the relative energy has been computed, at least for two reasons, one technical, one interpretative. From the technical viewpont, in fact, the generator of time translations is not a Killing vector for the dynamical metric nor for the background, and Dirichlet boundary conditions are not satisfied. We stress however that these conditions are not necessary for the conservation of the energy: in our case, in fact, they are not fulfilled but the current \eqref{corrente_testo} is still conserved on-shell and the integral of the relative superpotential \eqref{relsup} is strongly conserved.
Let us moreover remark that for homogeneous and isotropic universes it is not possible to ask for a matching of the dynamical metric with the background on $\D D$ since such a requirement would imply $g=\bar{g}$. Nevertheless the N\"other current can be computed with two non matching $g$ and $\bar{g}$ and it is conserved. The unfulfillment of Dirichlet boundary conditions does not 
affect the variation of the superpotential \eqref{relsup} that again provides \eqref{variation}.
The covariantly conserved quantities associated to the generators of the Poincar\'e subalgebra are moreover conserved also {\it in time} (according to the definition of Appendix \ref{intime}).

In the previously known standard examples the matching on the boundary, or the Killing property for the symmetry generator, were always explicitly asked for in order to ensure the vanishing of the integral \eqref{bordo}. This is, in our experience, the first example where both requirements are not satisfied still giving rise to time conservation (for all the symmetry generators considered).

From the viewpoint of physical interpretation we have to emphasize that, as was already pointed out by \cite{cooperstock}, isotropic universes are not asymptotically flat, and there is no way of going \<< infinitely'' far away from the sources to move a the test particle and check if our definition of \<<energy'' is compliant  with intuition. In any case our definition provides a covariantly conserved quantity associated to time translations that is also conserved {\it in time} and that certainly has a physical meaning.

Let us also remark that even if our starting point in deriving the superpotential \eqref{relsup} was to consider a $1$-parameter family of solutions $\{g_s\}$ connecting $g$ and $\bar{g}$, nothing prevent us from considering from the very beginning the Lagrangian \eqref{aug_lag_testo} with two solutions that are not homotopic (of course in this case it is no longer possible to study the variation of its superpotential, and the covariant ADM formalism is not any longer a valid motivation, but let us go on and se what comes out). In the case of two homogeneous ad isotropic cosmological solutions with different curvature (e.g., $-1$ for the dynamical solution, $0$ for the background) we get that the relative energy inside the surface $t=\text{const}$, $r=\text{const}$ is 
\[E_{(t,r)}^{1,-1}=  -\frac{ \bar{a}^3(t) r^3}{2  a^2(t)} \]
and considering the limit for $r \rightarrow \infty$ we get an infinite relative energy between the two configurations. This result is not surprising since to switch from a universe with flat  hypersurfaces of homogeneity to another one in which they have negative curvature, a change in the topology is needed and it seems reasonable that one needs to spend an infinite energy to perform such a task. We stress that we do not have a proof of the fact that energy diverges if and only if $g$ and $\bar{g}$ are not homotopic though to the best of our knowledge the only cases in which the energy is known to diverge are those when $g$ and $\bar{g}$ are not homotopic.

\section{Conclusions}
We applied variational calculus to calculate the relative covariantly conserved quantities between two homogeneous and isotropic cosmological models.
We found that if the two models share the same normalized spatial curvature their relative conserved quantities relative to an entire subalgebra of vector fields isomorphic to that of the Poincar\'e group are vanishing. In particular their relative energy, defined as the conserved quantity associated to time translations for a comoving observer, is vanishing.  We also found that it is possible to chose a vector field such as that the associated conserved quantity is both non vanishing and conserved {\it in time}. 

Now that we extended the prescription of Ref. \cite{augmented} to constrained systems (and in particular to relativistic fluids), further investigations will be devoted to describe conserved quantities of other systems  of the same kind such as Tolman-Bondi collapse solutions.

\appendix

\section{Conservation laws \<<in time''}\label{intime}
In the text we illustrated what we mean when saying that a quantity is covariantly conserved. In the literature, however, the word {\it conserved} is often associated to quantities that do not change in time. This is a different property and it is not manifestly covariant. However let us explain in which sense some covariantly conserved quantities are also conserved {\it in time}.
Let us consider a spacelike $3$-dimensional hypersurface $\Sigma$ and a timelike vector field $Y$. Let us assume that the flow of $Y$ drags the surface $\Sigma$ generating an ADM foliation $\Sigma_s$ where $s$ is the {\it time} associated to it at least locally. One can consider the 4-dimentional region $D$ spanned by the surfaces $\Sigma_{s}$ when the parameter runs from $s_0$ to $s_1$. It is bounded by the three surfaces $\Sigma_{s_0}$,  $\Sigma_{s_1}$ and $B=\cup_s \D \Sigma_s$.
Let us then consider a solution $y$ of field equations, a symmetry $\xi$ of the Lagrangian $L$ and the conserved current $\mathcal{E}(L,\xi,y)$. The covariantly conserved quantity $Q(L,\xi, \Sigma_{s_0}, y)$ obtained by integrating $\mathcal{E}(L,\xi,y)$ along $\Sigma_{s_0}$ is also conserved in time $s$ if such an integral does not depend on which of the surfaces $\Sigma_s$ the integral is computed on.
Being $\D D=  \Sigma_{s_0}\cup\Sigma_{s_1} \cup B$ one has
\begin{equation}
0=\int_D d \mathcal{E}(L,\xi,y)= \int_{\D D} \mathcal{E}(L,\xi,y)
=  \int_{\Sigma_{t_1}} \mathcal{E}(L,\xi,y) -\int_{\Sigma_{t_0}} \mathcal{E}(L,\xi,y) +\int_{B} \mathcal{E}(L,\xi,y) 
\end{equation}
and thus conservation in time is equivalent to the vanishing of the flow of the current along $B$:
\begin{equation} \int_{B} \mathcal{E}(L,\xi,y) =0. \label{bordo}\end{equation}
Different sets of sufficient conditions can be given in order to ensure the conservation in time of a covariantly conserved quantity. We do not list them here and we refer the reader to Ref. \cite{brownyork}.

\def\cprime{$'$}

\end{document}